\documentclass[conference]{IEEEtran} 
\IEEEoverridecommandlockouts
\usepackage{cite}
\usepackage{amsmath,amssymb,amsfonts}
\usepackage{algorithmic}
\usepackage{graphicx}
\usepackage{textcomp}
\usepackage{xcolor}
\usepackage[boxed,ruled,vlined,linesnumbered]{algorithm2e}

\usepackage{float}

\def\BibTeX{{\rm B\kern-.05em{\sc i\kern-.025em b}\kern-.08em
    T\kern-.1667em\lower.7ex\hbox{E}\kern-.125emX}}
    
\newcounter{Remark}
\newenvironment{Remark}
{
	\refstepcounter{Remark}
	\textbf{Remark \theRemark:}
}

\begin{document}

\title{Proof-of-Contribution-Based Design for Collaborative Machine Learning on Blockchain\\
}

\author{%
}
\author{Baturalp Buyukates$^{*1}$ \qquad Chaoyang He$^{*2}$  \qquad Shanshan Han$^{3}$ \qquad Zhiyong Fang$^{4}$ \\ Yupeng Zhang$^{4}$ \qquad Jieyi Long$^{5}$ \qquad Ali Farahanchi$^{6}$ \qquad Salman Avestimehr$^{1,2}$ \\ 
\normalsize \textsuperscript{1}Department of Electrical and Computer Engineering, University of Southern California, CA, USA \\
\normalsize \textsuperscript{2}FedML Inc., CA, USA \\
\normalsize \textsuperscript{3}Department of Computer Science, University of California Irvine, CA, USA \\
\normalsize \textsuperscript{4}Department of Computer Science and Engineering, Texas A\&M University, TX, USA \\
\normalsize \textsuperscript{5}Theta Labs Inc., CA, USA \\
\normalsize \textsuperscript{6}Camford Capital, CA, USA \\

\emph{buyukate@usc.edu} \qquad \emph{ch@fedml.ai} \qquad \emph{shanshan.han@uci.edu} \qquad \emph{zhiyong.fang.1997@tamu.edu} \\ \emph{zhangyp@tamu.edu} \qquad \emph{jieyi@thetalabs.org} \qquad \emph{ali@camford.vc} \qquad \emph{avestime@usc.edu} \thanks{* denotes equal contribution.}}

\maketitle

\begin{abstract}
We consider a \emph{project (model) owner} that would like to train a model by utilizing the local private data and compute power of interested \emph{data owners, i.e., trainers.} Our goal is to design a data marketplace for such decentralized collaborative/federated learning applications that simultaneously provides i) proof-of-contribution based reward allocation so that the trainers are compensated based on their contributions to the trained model; ii) privacy-preserving decentralized model training by avoiding any data movement from data owners; iii) robustness against malicious parties (e.g., trainers aiming to poison the model); iv) verifiability in the sense that the integrity, i.e., correctness, of all computations in the data market protocol including contribution assessment and outlier detection are verifiable through zero-knowledge proofs; and v) efficient and universal design. We propose a blockchain-based marketplace design to achieve all five objectives mentioned above. In our design, we utilize a distributed storage infrastructure and an aggregator aside from the project owner and the trainers. The aggregator is a processing node that performs certain computations, including assessing trainer contributions, removing outliers, and updating hyper-parameters. We execute the proposed data market through a blockchain smart contract. The deployed smart contract ensures that the project owner cannot evade payment, and honest trainers are rewarded based on their contributions at the end of training. Finally, we implement the building blocks of the proposed data market and demonstrate their applicability in practical scenarios through extensive experiments.
\end{abstract}


\IEEEpeerreviewmaketitle

\section{Introduction}

Over the past decade or so, our lives have transitioned into a fully connected reality where our smartphones, wearables, and other smart devices are constantly able to collect information about us and our surroundings. With this amount of available data, more-informed decisions (whether by smart assistants, service providers, vendors or government officials) can be made by leveraging machine learning (ML), in particular training of deep neural networks (DNNs).
For example, real-time location data from driver devices can help route traffic around the city, improving the congestion situation in rush hours; vendors can improve their advertisement as well as their line-up of offerings based on analyzing information from collective information from their clients' behaviors. Even though an enterprise may only be interested in learning useful aggregate information about its user population, data privacy has become an increasingly precious and critical commodity that individuals and/or enterprises are suspicious to give up, even for a greater-good goal. To deal with these challenges, collaborative/federated learning (FL) has been introduced to get benefit from outsourcing the computations as well as using participants' local datasets to learn a more accurate model \cite{mcmahan2017communication}.

\begin{figure}[t]
	\centering
    \includegraphics[width=0.3\textwidth]{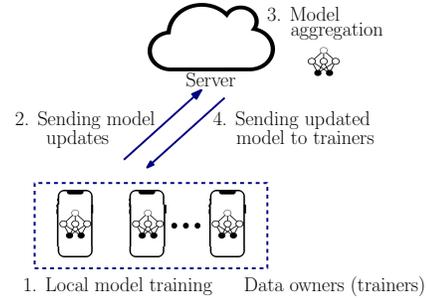}
	\caption{Federated learning enables machine learning over decentralized data. An FL system consists of a central server and a certain number of trainers.}
	\label{fig:fl}
	\vspace{-5mm}
\end{figure}

\noindent\textbf{Federated learning (FL)}.
As shown in Fig.~\ref{fig:fl}, an FL system consists of a central server (aggregator) and a certain number of data owners (trainers) that collaboratively train a global model under the orchestration of the aggregator.
Each data owner $i$ has a private dataset $D_i=\{(x_1,y_1),\dots,(x_{n_i},y_{n_i})\}$ of size $n_i$, $i \in [L]$. The goal of the aggregator is to minimize
\begin{align}
    L(\mathbf{w})=\sum_{i=1}^{L} \frac{n_i}{n}L_i(\mathbf{w}),
\end{align}
where $L_i(\mathbf{w})= \frac{1}{n_i} \sum_{(x,y) \in \mathcal{D}_i} \ell(\mathbf{w};(x,y))$ is a predefined empirical local loss function over the dataset $\mathcal{D}_i$.

In each round $t$, the aggregator sends the updated global model $\mathbf{w}{(t)}$ to the trainers. Each trainer $i$ computes gradient vector $\nabla_{\mathbf{w}}L_i(\mathbf{w})$ over its private dataset and computes a local model $\mathbf{w}_i(t)$. When the data owners have equal-sized datasets, the aggregator computes an updated global model with $\mathbf{w}{(t+1)} = \frac{1}{L}\sum_{i \in [L]} \mathbf{w}_i(t).$ Upon aggregation, the aggregator pushes $\mathbf{w}(t+1)$ back to the trainers and this iterative process continues until the model converges.

Over the past few years, there has been a surge of research in FL, in particular on algorithmic developments to address data heterogeneity~\cite{ghosh2019robust}, personalization~\cite{fallah2020personalized}, and fairness~\cite{li2019fair}; privacy enhancements to address potential data leakage from locally trained models of the participants~\cite{geyer2017differentially, Bonawitz17, so22lightsecagg, kadhe2020fastsecagg} and security enhancements to address potential attacks in FL systems~\cite{lyu2020privacy}, as well as applications of FL to various ML domains~\cite{li2020review}. 

Despite all these advances and benefits, collaborative ML systems (including FL) still need to provide sufficient incentive mechanisms to the participants. This is particularly important with the surge of Web3 era and rising interests for content and data monetization. To that end, data marketplaces can be utilized to incentivize interested parties to participate in the learning algorithm, use their local data, and perform heavy computations in exchange for monetary compensation.

In this work, our goal is to design a blockchain-empowered data marketplace for collaborative learning that is executed through a smart contract. We have a project (model) owner that wants to train a model by utilizing the compute power and private datasets of data owners (trainers)\footnote{Throughout the text, we use terms project owner and model owner interchangeably. Similarly, we use terms data owner and trainer interchangeably.}. Key features of our design are proof-of-contribution-based fair reward allocation to trainers according to their actual contributions to the model; privacy-preserving training such that private local datasets of the trainers never leave their devices; verifiability such that the integrity of the entire data market operation is verifiable; robustness against malicious parties; and efficient implementation and universal applicability to different learning tasks. We utilize a decentralized storage infrastructure so that large ML datasets and models are stored off-chain. Heavy computations such as model aggregation, outlier detection, and contribution assessment are performed off-chain by a processing node, which we call the aggregator. The use of blockchain guarantees that the project owner cannot evade payment.

We propose a new metric called the \emph{randomized leave-one-out (RLOO)} to assess the contribution of each trainer to the model. To detect malicious trainers, we propose a novel outlier detection scheme that is inspired by the three sigma rule \cite{three_sigma} and the Krum defense \cite{krum}. To make the entire data market operation verifiable, we use zero-knowledge proofs (ZKPs) for model aggregation, outlier detection, and contribution assessment, which are generated by the aggregator and verified by the smart contract. For ZKPs, we use the Groth16~\cite{groth2016size} zkSNARK scheme implemented in the Arkworks library~\cite{arkworks}. We implement the key components of the proposed design and show their performance in practical scenarios through experiments. We highlight that the proposed data market design is modular and each of the proof-of-concept building blocks (contribution assessment, outlier detection, and the ZKP protocols) can be modified with different trade-offs.

\noindent\textbf{Related works.}
Traditional data marketplaces were proposed for ML applications \cite{pei2020survey,chen2019towards,liu2021dealer} considering ML model pricing and/or compensations for data owners. In \cite{liu2021dealer}, data owners are compensated based on the quality of their data along with their privacy sensitivity. Unlike our work, in \cite{liu2021dealer}, not only \emph{raw} user datasets are observed by the data market but also all parties in the marketplace are assumed to be honest, i.e., there is no verification mechanism on the operation of the data market. Further, in such conventional designs, during data transactions, data market becomes a fully trusted and authoritative third-party \cite{chen2017bootstrapping}, which may be undesired in many practical collaborative ML scenarios as it constitutes a single point of failure. To resolve this, references \cite{chen2017bootstrapping,ozyilmaz2018idmob,zhang2015iot,worner2014your} focus on building decentralized data marketplaces but fail to address potential data/model privacy issues. Focusing on privacy, in \cite{duan2019aggregating,niu2018unlocking,koutsos2021agora} blockchain-based data market schemes are implemented such that encrypted data are uploaded on the blockchain network. The problem here is two-fold. First, there is a memory limitation at each block so that the blockchain cannot support large ML models. Second, when the uploaded data are encoded, the system can only compute simple operations (such as linear combination of data) and cannot generalize to more complex and practical ML computations. FL in decentralized frameworks such as blockchain is studied in \cite{kim2019blockchained, zhao2020privacy, lyu2020towards,ma2021transparent, liu2020fedcoin, liang2021omnilytics}. Reference \cite{kim2019blockchained} discusses a serverless blockchained FL system, called \emph{BlockFL}, where the role of the server in FL (i.e., gathering local updates and model aggregation) is performed by the blockchain network. Their proposed approach rewards the participants based on their dataset size. In \cite{zhao2020privacy}, a blockchain-based FL system is proposed that uses differential privacy to protect data privacy. 
The reward mechanism in \cite{zhao2020privacy} is distance-based which may not be suitable for non-i.i.d.~data distributions among the data owners. A contribution-based FL system on blockchain is proposed in \cite{lyu2020towards} by only considering i.i.d.~data distribution. 
Many existing works in collaborative/federated learning \cite{liu2021dealer, ma2021transparent, liu2020fedcoin} use Shapley value (SV) associated with each party to evaluate the contributions of the participants. In \cite{ma2021transparent}, authors propose a group-based SV approach that is suitable with secure aggregation techniques for contribution evaluation. In \cite{liu2020fedcoin}, authors design a blockchain-based peer-to-peer payment system called \emph{FedCoin} that rewards participants based on their SVs. We want to highlight that in practical collaborative ML systems, computing the SV is not feasible when the number of data owners is large. Recently, \cite{liang2021omnilytics} proposed a blockchain-based data market named \emph{OmniLytics} that gives the same reward to all honest data owners, which (albeit being efficient) has fairness implications and is prone to alterations by adversarial participants.

\noindent \textbf{Limitations of the existing data market designs.} Existing data market designs for collaborative/federated learning suffer to simultaneously provide: i) contribution-based fair rewards, ii) privacy, iii) robustness against malicious parties, iv) verifiability, v) efficiency and universal applicability to heterogeneous data distributions and different learning tasks, all of which motivate us to propose a novel blockchain-based data marketplace for collaborative ML.
\section{Problem Setting and Contributions}

\begin{figure}[t]
	\centering  \includegraphics[width=0.49\textwidth]{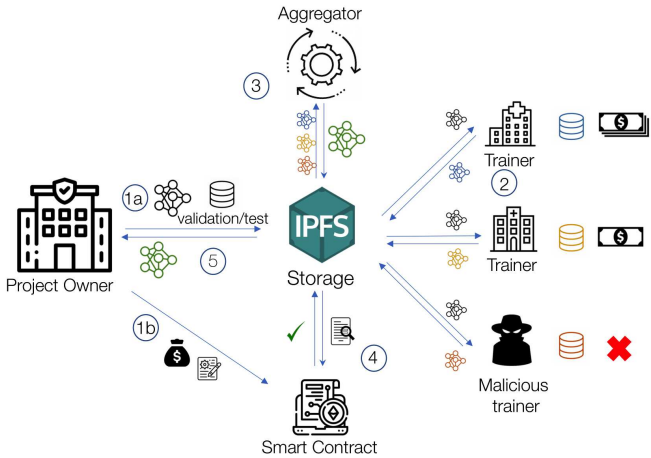}
	\caption{The proposed data marketplace design for collaborative/federated learning on blockchain.} 
	\label{fig:DMA}
    \vspace{-5mm}
\end{figure}

\noindent \textbf{Goals.}
Our goal is to design a blockchain-empowered data marketplace in which i) the data owners get rewarded fairly based on the contribution of their locally trained models, ii) any data movement from the data owners is not allowed and no other party can see the local datasets of the data owners, iii) robustness is provided against adversarial behavior of the malicious participants, iv) all of the computations including model aggregation, contribution assessment, and outlier detection are verifiable, and v) the implementation is efficient and applicable to practical scenarios for different learning tasks under varying degrees of data heterogeneity.

\noindent \textbf{Proposed Data Market Overview and Contributions.} We design a trustworthy proof-of-contribution-based data marketplace for collaborative ML on blockchain that satisfy all these goals mentioned above (see Fig.~\ref{fig:DMA}). The proposed system consists of a project owner, data owners (trainers), an aggregator, and a distributed storage infrastructure. We use a decentralized P2P storage infrastructure called the Interplanetary File System (IPFS) as in \cite{Passerat19}. For each stored file, IPFS generates a unique content identifier (CID), which points at the content itself (as opposed to its location in centralized web) that is used to access the file.  

The project (model) owner would like to crowdsource the training of its model to interested data owners, i.e., trainers, in exchange for monetary returns. The data market commences with the project owner creating a smart contract by depositing some tokens and task descriptions, i.e., CIDs of validation/test datasets (which are already uploaded to IPFS), model description, training hyperparameters to initialize the project. With the execution of the smart contract on blockchain, interested data owners join the system. Once sufficiently many trainers join the system, training starts by the aggregator initializing the model and storing the base model in IPFS. Each participating trainer pulls the current global model from the storage, performs local training, and uploads its local model to IPFS. The aggregator retrieves the corresponding local models and performs outlier detection/removal and weighted aggregation of the accepted honest local models as well as an evaluation of the aggregated model on the validation dataset (for hyperparameter tuning). The aggregator then submits the benign data owner IDs (addresses) and zero-knowledge proofs (ZKPs) for outlier detection and the model aggregation results to the smart contract, which are then verified by the smart contract. If the verification is successful, the aggregator is notified to start the next round of training. Otherwise, the training run is terminated. The system continues for a certain number of rounds or until the aggregator determines that the performance is saturated.
 
Once the training is completed, the aggregator performs contribution assessment of the trainers\footnote{Contribution assessment is performed by the aggregator at the end of training using the best training run.} and generates ZKPs for the contribution assessment as well as the per round validation accuracy computations performed during training. These proofs and the contribution vector are submitted to the smart contract, which then performs the verification. If the contribution assessment and per round validation accuracies are verified, the rewards are distributed from the project owner to the participating trainers based on the contribution of each trainer. In distributing the rewards, we give a fixed fee per round to each participating honest trainer in addition to the contribution-based payment. In our design, the exact amount of total reward (to be distributed from the tokens deposited by the project owner) is based on the accuracy of the final model. 
 
The key features of the proposed data market are as follows: 
\begin{enumerate}
    \item \textbf{Contribution-based reward allocation mechanism:} The total reward is based on the accuracy of the final model whereas the individual rewards of each trainer is based on how much each of them contributes to the overall performance of the final model. Thus, not only the project owner has control over the total reward payment but also trainers are compensated fairly. 
    
 	
    \item \textbf{Privacy:} In our marketplace, no party learns about the private datasets of the data owners since datasets are not required to be moved from the data owner (trainer) side. Since the local models stored at IPFS are encrypted, only the aggregator can see the individual models of the trainers. Further, trainers may opt for differential privacy \cite{wei2020federated} in uploading their local models to the storage. 
    
    \item \textbf{Robustness:} The use of blockchain guarantees that the project owner cannot evade payment after training. Thanks to the employed verification mechanisms, at each step of the data market execution, the proposed design is robust against malicious parties including but not limited to i) malicious trainers poisoning their models to disrupt the convergence of the global model or make the final result incorrect, ii) a malicious aggregator uploading incorrect aggregation, outlier detection, and/or contribution assessment results, and iii) a malicious project owner that wants to evade payment. 
  	
    \item \textbf{Verifiability:} The integrity of the model aggregation, contribution assessment, and outlier detection computations are verified via ZKPs generated by the aggregator.
    
    \item \textbf{Efficiency and universal applicability:} The proposed scheme can be implemented with the existing tools such as ZKPs for matrix multiplication and blockchain. Unlike most of the prior blockchain-based works (e.g.,\cite{ma2021transparent,liang2021omnilytics,zhao2020privacy,lyu2020towards}), in our design, computations are not repeated by the miners for verification purposes. In our design, just the aggregator performs the computations off-chain and the smart contract verifies the results without doing heavy computations on-chain, which in turn makes our design lightweight and efficient in implementation. In addition, the proposed data market has a modular design and is applicable to practical scenarios with data heterogeneity and trainer dropouts. Our design can be utilized for various learning tasks as we do not implement any encryption on the trainer data. 
\end{enumerate}

%

\section{Data Market Design: Building Blocks}

\subsection{Proposed Contribution Assessment Metric }\label{subsection:contributionassessment}

To have a fair resource/credit/reward allocation \cite{liu2020fedcoin,song2019profit} in collaborative ML, one needs to assess the contribution of each local model. There are various methods for computing the contribution of the locally trained models in the literature \cite{song2019profit,wang2019interpret,wang2019measure}. In \cite{wang2019interpret}, Shapley value (SV) \cite{roth1988shapley} associated with each local model is used as a contribution assessment metric. The main idea of SV is to compute the marginal utility of each model (i.e., each trainer) considering different trainer permutations. In collaborative ML, this utility function is the test accuracy. The main disadvantage of SV is that one needs to retrain the model for each permutation to compute the exact SV, which incurs a significant computation load (that is exponential in the number of federated trainers), thus making exact SV impractical in collaborative ML as a metric for contribution assessment.

Efficient approximations of the SV method has been studied in the literature \cite{song2019profit, Wang2020, Liu2022gtg}. Among these, the current state-of-the-art SV approximation technique is GTG-Shapley \cite{Liu2022gtg}. In GTG-Shapley, local model updates are used to reconstruct the corresponding model under different trainer permutations (as opposed to retraining from scratch) and certain within round and between round truncation techniques are implemented to speed up the computation. 

\noindent\textbf{Randomized leave-one-out (RLOO).} In our data market design, as mentioned earlier, the aggregator performs the contribution evaluation of the trainers. Considering the verifiability of this contribution evaluation, the smart contract needs to receive proofs for each accuracy evaluation performed by the aggregator (as detailed in Section~\ref{subsection:ZKP}), in which case the number of permutations used for SV approximation burdens the system performance. Motivated by this, in our data market design, we propose an even more efficient randomized leave-one-out (RLOO) method (compared to GTG-Shapley) to assess the contribution of each local model. Inspired by the data point deletion method of \cite{wang2019measure}, in RLOO, by removing one model from the aggregate at a time, we measure how the accuracy of the aggregate model changes to deduce the affect of a particular local model. In RLOO, instead of using the aggregate of all local models, the aggregator first samples a subset of the trainers and uses the aggregate of the local models of the sampled subset, hence the name randomized leave-one-out.

Let $N^{(t)}$ denote the number of honest trainers in training round $t$. To compute the contribution of trainer $i$ in round $t$, denoted by $C^{(t)}_i$, the aggregator first samples a $\log N^{(t)}$ subset of trainers (making sure that trainer $i$ is in the subset) and computes their aggregate model $\mathbf{w}^{(t)}$. Then
\begin{align}\label{approx_sv}
	C^{(t)}_i=\text{Acc}\left(\mathbf{w}^{(t)}\right)-\text{Acc}\left(\mathbf{w}^{(t)}_{-i}\right),
\end{align}
where, $\mathbf{w}^{(t)}_{-i}$ denotes the aggregate model of the subset except 
the local model of trainer $i$. In (\ref{approx_sv}), $\text{Acc}(\mathbf{x})$ is the accuracy function of model $\mathbf{x}$ over the test data. If (\ref{approx_sv}) is a positive number, we say the $i$th data owner contributes to the model. When (\ref{approx_sv}) is negative, this indicates that the global model is better off without the local model of the $i$th data owner in that training round. Finally, we normalize the contribution values $C^{(t)}_i$ of each honest data owner $i$ in each round $t$ and report the sums over the entire training run as total individual contribution values $C_i$ of the honest data owners.


\subsection{Proposed Outlier Detection Mechanism}
\label{subsection:outlierdet}
Malicious trainers or outside attackers in collaborative ML can harm training by employing Byzantine attacks~\cite{fang2020local, chen2017distributed} or backdoor attacks~\cite{bagdasaryan2020backdoor,wang2020attack}. Byzantine attacks aim to disrupt the convergence of the global model, while backdoor attacks aim to mis-classify some data to a wrong class via backdoors. Existing literature on Byzantine-resilient learning and mitigation of adversarial behavior includes \cite{krum,yang2019byzantine,chen2017distributed,fung2018mitigating}. These existing approaches either require knowing the number of malicious trainers in advance~\cite{krum} or re-weighting trainer submissions~\cite{fung2018mitigating}, which are not practical in real systems, since an adversary will not notify the system before attacking, and re-weighting will tamper the aggregation results if there is no attack.
A successful outlier detection approach should simultaneously satisfy the following: 1) the approach should detect whether an attack has happened; 2) if that is the case, the approach needs to handle the poisoned models to mitigate their effects; and 3) the approach should not harm benign trainer submissions when handling outliers.

In this work, we propose a novel two-stage outlier detection mechanism satisfying all three properties. The proposed approach first performs a cross-round check to detect whether any attacks happened. In case of attacks, it then performs a cross-trainer check to remove malicious trainer submissions. 


\noindent\textbf{Cross-round check.}
The aggregator computes the cosine similarity between the local models submitted by each trainer in the current round and the previous round. Specifically, for trainer $i$ and the local model $\mathbf{w}_i^{(t)}$ in round $t$, it computes
\begin{align}\label{cosinesim}
    S_c(\mathbf{w}_i^{(t)}, \mathbf{w}_i^{(t+1)}) = \frac{\mathbf{w}_i^{(t)}\cdot\mathbf{w}_i^{(t+1)}}{\lVert\mathbf{w}_i^{(t)}\rVert \lVert\mathbf{w}_i^{(t+1)}\rVert}.
\end{align}

Any cosine score $S_c$ that is lower than a threshold $\gamma$ indicates that an attack happened in the current round. 

\noindent\textbf{Cross-trainer check. }
In case of attacks, the aggregator performs a cross-trainer check to remove malicious models. The cross-trainer check inherits the ideas of the three sigma rule \cite{three_sigma} and the Krum algorithm~\cite{krum}. The three sigma rule indicates that in normal distributions the percentage of values within one, two, and three standard deviations of the mean are 68\%, 95\%, and 99.7\%, respectively. Enabled by the central limit theorem~\cite{clt} and certain transformations \cite{aoki1950stability, osborne2010improving,sakia1992box, weisberg2001yeo}, the three sigma rule has many real-world applications and has previously been used in outlier detection~\cite{han2019iterative}.

The main idea of the cross-trainer check is to compute the distance between each trainer model and the average trainer model. Using these distances, the aggregator computes a boundary for outlier detection and removes trainer models which lie outside this boundary. 

In particular, the aggregator computes Euclidean distances $\lVert\mathbf{w}_i-\mathbf{w}_{\mathit{avg}}\rVert$, where $\mathbf{w}_i$ denotes the $i$th trainer model, $i \in [L]$, and $\mathbf{w}_{\mathit{avg}}$ is the average trainer model, computed as a ``middle point". In the first round of training, $\mathbf{w}_{\mathit{avg}}$ is computed using the m-Krum algorithm~\cite{krum} by taking $m=\frac{L}{2}$, when the aggregator receives $L$ locally trained models, assuming we have honest majority. In other training rounds, the average is computed by simply averaging all benign trainer models.

We let $\mathcal{S}_E$ denote the list of the Euclidean distance scores with cardinality $|\mathcal{S}_E|$. The aggregator then computes an approximate normal distribution $N(\mu, \sigma)$ using $\mathcal{S}_E$ with
\begin{align}
    \mu = \frac{\sum_{\ell\in \mathcal{S}_E} \ell}{|\mathcal{S}_E|}, \hspace{2em}
    \sigma = \sqrt{\frac{\sum_{\ell\in \mathcal{S}_E} (\ell-\mu)^2}{|\mathcal{S}_E|-1}}.
\end{align}
Invoking the three sigma rule, the aggregator computes the outlier boundary as $\mu + \sigma$ so that only the trainer models 
with scores less than $\mu + \sigma$ are kept, i.e., the ones that are at most one standard deviation away from the mean. The others are deemed as outliers and removed (see Algorithm~\ref{alg: outlier_detection}).

\begin{algorithm}[t]
	\caption{Cross-trainer check in outlier detection}
	\label{alg: outlier_detection}
	\textbf{Inputs:} 
 {$\tau$: training round; 
 $\mathcal{W}_{\tau}$: trainer models in round $\tau$, $|\mathcal{W}_{\tau}| = L$} 

\nl{\bf function $\boldsymbol{\mathit{Detect}\_\mathit{Outlier}( \mathcal{W}_{\tau}, \tau)}$\nllabel{ln:detect_outlier}}
\Begin{

\nl \If{$\tau = 1$} {
 {// Use m-Krum \cite{krum} with $m=L/2$ assuming honest majority }
 
$\textbf{w}_{\text{avg}}\leftarrow\mathit{compute}\_\mathit{avg}\_\mathit{model}\_\mathit{with}\_\mathit{krum}(\mathcal{W}_{\tau})$
} 

{\nl {// Compute L2 distances for each trainer model}}

\nl  $\mathcal{S}_E \leftarrow \mathit{compute}\_\mathit{L2}\_\mathit{scores}(\mathcal{W_{\tau}}, \textbf{w}_{\text{avg}})$ 

{\nl {// Compute $\mathcal{N}(\mu, \sigma)$ using the L2 scores}}

{\nl $\mu \leftarrow \frac{\sum_{\ell\in \mathcal{S}_E} \ell}{|\mathcal{S}_E|}$, $\sigma \leftarrow \sqrt{\frac{\sum_{\ell\in \mathcal{S}_E} (\ell-\mu)^2}{|\mathcal{S}_E|-1}}.$}

\nl 
\For{  $i \leq |\mathcal{S}_E|$}{

\nl \If{$\mathcal{S}_E[i] > \mu + \sigma$}{ remove $\mathbf{w}_i$ from $\mathcal{W}_{\tau}$
}
}

\nl {
$\textbf{w}_{\text{avg}}\leftarrow \mathit{average}(\mathcal{W}_{\tau})$
}

\nl\textbf{return} $\mathcal{W}_{\tau}$
}
\end{algorithm}

\subsection{ZKPs for Aggregator Computations}
\label{subsection:ZKP}

Introduced in \cite{Goldwasser19}, zero-knowledge proofs (ZKPs) enable a prover to convince a verifier that a computation on the prover's secret input (called the witness) is correct via a short proof, without revealing any information about the secret input. ZKPs guarantee that the prover cannot cheat (integrity property) as well as that the prover's secret input remains confidential (zero-knowledge property). Due to their strong integrity and privacy guarantees, ZKPs have been utilized in blockchain systems \cite{sasson2014zerocash}, and machine learning applications \cite{lee2020vcnn, feng2021zen, liu2021zkcnn}.

In the proposed data market design, heavy computations such as outlier detection, model aggregation, and contribution assessment are performed off-chain by the aggregator. Since the aggregator is essentially an untrusted server, it may send inaccurate/incorrect results in order to save its computation time and resources or due to malicious behavior. Thus, the smart contract requires proof of the aggregator's computations. In our work, we utilize zero-knowledge Succinct Non-interactive Arguments of Knowledge (zkSNARKs). zkSNARKs offer proofs that are short in length and easily verifiable without interactions between the prover and verifier, thereby are desirable for blockchain systems in which the verifier time/resources are critical. In the subsequent text, the prover is the aggregator and the verifier is the smart contract.

\noindent{\textbf{ZKP-compatible language.}} 
The first challenge of applying ZKP protocols is to express the computations in ZKP-compatible languages. ZKP protocols model the computation as an arithmetic circuit with modular addition and multiplication gates over a prime field. However, the computations in our data market design are on real numbers, such as the outlier detection and the machine learning computations. To address this issue, in our system, we build ZKP schemes using the Arkworks library~\cite{arkworks}, which compiles the description of an arithmetic circuit in a front-end language similar to Rust to the back-end of ZKP protocol automatically. On top of the library, we implement a class of fixed-point numbers and operations. We also develop optimized gadgets for common computations used in our data market design, including the ReLU function, maximum and square-root. Compared to naive implementations of these functions as circuits, the number of gates is reduced significantly.

\noindent{\textbf{ZKP for outlier detection.}}
A zero-knowledge proof for matrix multiplication constitutes the basis of the verification schemes we use for the outlier detection and the model aggregation in the proposed data marketplace design. To verify a matrix multiplication $AB=C$, where $A,B,C$ are of size $n\times n$, instead of proving the computation step-by-step with $O(n^3)$ multiplication gates, we apply the Freivalds' algorithm~\cite{freivalds1977probabilistic} to test the correctness of the result with a random vector. The verifier generates a random vector $v$ of length $n$, and then checks $A(Bv)\stackrel{?}{=}Cv$ inside the ZKP scheme. The size of the circuit is reduced to $O(n^2)$ from $O(n^3)$ in the naive solution. This scheme requires a commit-and-prove zkSNARK~\cite{campanelli2019legosnark} and the random vector is generated after the prover commits to the result of the matrix multiplication. 

We utilize the idea of verifying the computation again to design an efficient protocol for square-root, which is used in Algorithm~\ref{alg: outlier_detection}. To verify that $x = \sqrt{y}$ is computed correctly, we ask the prover to provide the result $x$ and check in the ZKP that $x^2$ is close to $y$. Note that as the ZKP works over a prime field, it is not correct to check that $x^2 = y \mod$ $p$. Instead, we represent $x$ and $y$ as fixed-point numbers using our new class and check that $x^2 \le y$ and $(x+1)^2 \ge y$, where the ``+1'' denotes the adding one to the least significant bit of the fixed-point number represented in the prime field as an integer. In this way, the cost of a square-root computation is improved to two multiplications and two comparisons.


\noindent{\textbf{ZKP for model aggregation.}}
The model aggregation takes the average of all the local models, which requires a division by the number of accepted (benign) trainers. To verify a fixed-point division $c = a/b$, we ask the prover to provide the quotient $c$ and the remainder $r$, and check that the $a = b\cdot c+r$ and that the remainder $r$ is less than the divisor $b$. 

In model aggregation, we further observe that if all trainers behave honestly, the divisor is the number of trainers, a publicly known value. Thus, we ask all the trainers to divide their local models by the number of trainers before submitting them to the aggregator, which saves the cost of division completely. The trainers will tend to do so honestly since otherwise they can be detected as outliers or have a lower reward.

\noindent{\textbf{ZKPs for contribution assessment.}} \label{subsection:ZKP_ML} In order to verify the contribution assessments performed by the aggregator, we utilize ZKPs for the accuracy evaluation of an ML model. Here, the goal is to prove the accuracy of a secret model on a committed dataset. We utilize the ZKP for matrix multiplications described above for the linear layer(s) of a logistic regression and a neural network to improve the efficiency significantly. In addition, we avoid computing the costly sigmoid and softmax functions in the last layer and instead select the maximum value from the vector directly. This does not change the result of the inference and the accuracy of the ML model. 

\noindent{\textbf{Choice of the ZKP system.}} In our implementation, we use the Groth16~\cite{groth2016size} zkSNARK scheme implemented in the Arkworks library~\cite{arkworks} for all the computations described above. The main reason is that the scheme has the smallest proof size (128 bytes) and can be verified directly in Ethereum's smart contracts with native instructions. There are recent papers on more efficient ZKPs for ML inferences and accuracy~\cite{liu2021zkcnn}, but the proof there is too big to be posted on-chain and verified by a smart contract (hundreds of KBs). Achieving both efficient prover time and short proof size is left as an interesting future work. Our ZKP implementation is a proof-of-concept building block for the entire design of the data market, and we can switch to other existing ZKP schemes to implement the same computations with different trade-offs. 

\section{Data Market Design: Detailed Operation}
\label{section:proposedscheme}

In this section, we present the proposed blockchain-based data market for collaborative ML in more detail. 
We have a single project (model) owner, data owners (trainers), and an aggregator. The project owner initiates the data market protocol by deploying a smart contract on the blockchain. Then interested data owners join the system incentivized by the monetary returns offered in the contract by the project owner. The aggregator is a processing node that performs certain computations including assessing trainer contributions, removing outliers, updating hyperparameters, and so on. 

 
In conventional schemes such as OmniLytics \cite{liang2021omnilytics}, the project owner uploads all the data as well as the initial model to blockchain. However, due to memory limitations, it is not possible to put all these on the blockchain network. That is why, in our design, we utilize a decentralized storage structure such as IPFS. Once a piece of information is stored on IPFS, a unique content identifier (CID), which is essentially a hash value, is provided by the storage. That CID is later on used to access that particular information. By utilizing IPFS, the project owner (and any other entity of the proposed data market for that matter) avoids storing the model, datasets or any other large data on the smart contract (on-chain). Instead, we only store the CIDs on the smart contract, overcoming any memory-related issues on blockchain. This is one of the main differences between the proposed data market and the conventional data marketplace schemes. 

\begin{figure}[t]
	\centering  \includegraphics[width=0.49\textwidth]{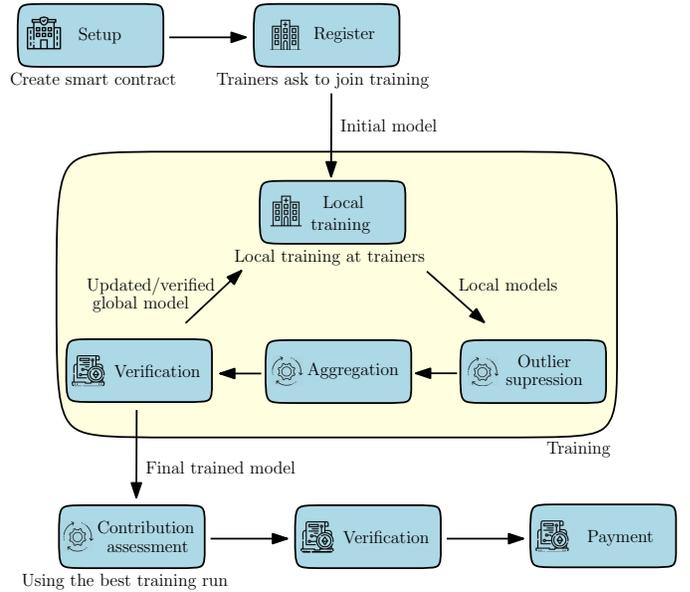}
	\caption{Flowchart of the proposed design executed through a smart contract. In each box, the icon (see Fig.~\ref{fig:DMA}) shows which party performs that action.}
	\label{fig:state}
    \vspace{-5mm}
\end{figure}
The execution of the proposed data marketplace is as follows (see Fig.~\ref{fig:state} for a flowchart of our design).


\subsection{Initialization}
\label{subsection:initialization}

\noindent{\textbf{Start.}} In order to start a data market session, the project owner uploads its validation and test datasets to IPFS and deposits some tokens as well as creates a smart contract that includes the task descriptions for the learning task. These descriptions include the CIDs of the validation and test datasets, learning model description, training hyperparameters, and so on. The deployed smart contract specifies the total amount of reward that will be available to the trainers based on the performance of the final trained model on the test data. 

\noindent{\textbf{Registration.}} With the execution of the smart contract on blockchain, the contract waits until $L$ data owners register for training the project owner's model on their local data. Each participating data owner provides two important arguments: a public key that is used to perform asymmetric encryption for secure IPFS storage and P2P communication along with a blockchain account address to be used to distribute rewards (tokens) based on proof-of-contribution after training. Once $L$ data owners join, the contract moves to the training phase.

\begin{Remark}
    In this work, our focus is on the trustworthy, verifiable operation and contribution-based rewards to the participating trainers. A separate open problem is selecting the trainers for the learning task at hand. For this, we envision that the smart contract can check the validity, i.e., whether that trainer has enough number of samples and good computational resources, and legality, i.e., good reputation, of each interested trainer. Trainers that do not meet the criteria in the smart contract would be rejected. How to select the best trainers in a verifiable/privacy-preserving manner in collaborative ML is beyond the scope of this work.
\end{Remark}

\subsection{Training Phase - Common Operations Per Round}
\label{subsection:perroundoperations}
Training starts by the aggregator initializing the model weights and storing the base model in IPFS. The aggregator records the CID of the base model to the smart contract.

\noindent{\textbf{Local training.}} Registered trainers pull the base model from the storage using its CID and start local training per the project owner’s specifications. Once the local training finishes, each trainer encrypts and uploads its local model to IPFS and records the corresponding CID to the smart contract. Thanks to this encryption, only the aggregator who knows the shared secret can decrypt the local model of each trainer. At this stage, trainers can further protect the privacy of their models by adding DP noise \cite{wei2020federated}. There is a trade-off between obtaining rewards and privacy, i.e., if a trainer adds more noise, then the performance of the local model reduces, thereby possibly decreasing that trainer's monetary return from training.

\noindent{\textbf{Outlier detection.}} When the locally trained models are uploaded by the trainers, the aggregator downloads the local models and decrypts each of them by using a symmetric Diffie-Helmann (DH) secret derived from the trainer's public key and the aggregator's private key. Upon decrypting the local models, the aggregator performs outlier detection as described in Section~\ref{subsection:outlierdet}. Next, the aggregator generates a proof for its outlier detection computation as described in Section~\ref{subsection:ZKP}. Once the outlier detection is performed, the aggregator records the benign trainer IDs (addresses) as well as the associated proof on the smart contract. The IDs of the benign trainers can later be utilized to adjust the reputation of the trainers.

\noindent{\textbf{Model aggregation.}} After removing outliers, the aggregator performs weighted aggregation of the honest models. Once the aggregated model is computed, the aggregator evaluates it on the validation dataset to guide the trainers to modify/tune the training hyperparameters. If the accuracy is not as desired, the aggregator can terminate that particular training run early. Upon aggregation, the aggregator writes the updated model to IPFS and uploads the corresponding CID, possibly modified hyperparameters, and a ZKP for the model aggregation (as described in Section~\ref{subsection:ZKP}) to the smart contract. 

\noindent{\textbf{Verification of outlier detection and model aggregation.}} The smart contract uses the corresponding ZKPs to verify the integrity of the outlier detection and the model aggregation results. If the verification is successful, the aggregator is notified to start the next training round. In the next round, each trainer uses the updated model and hyperparameters for local training, and the above steps are repeated for a certain number of rounds or until the aggregator determines that the performance is saturated.

\subsection{Training Phase - Final Round}
\label{subsection:finalround}

When the training is completed, the project owner releases the test dataset stored on IPFS to the aggregator so that the aggregator computes the performance of the final trained model. Using this, the smart contract determines the total contract reward. Here, the exact amount of the total reward to be distributed is based on the improvement in the project owner's model and is specified in the smart contract.

\noindent{\textbf{Contribution assessment.}} The aggregator evaluates the contribution of data owners in each training round as in Section~\ref{subsection:contributionassessment}. It performs this contribution evaluation using the test dataset at the end of the training (as opposed to at the end of each round) as it uses the best training run to compute contributions (in order not to perform any unnecessary heavy computations on the unused training runs that yield worse models). Once the contribution vector of the honest trainers $\mathbf{C}=[C_1,C_2,\dots,C_{L'}]$ is determined, the aggregator submits it to the smart contract along with a proof as described in Section~\ref{subsection:ZKP_ML}. Here, $C_i$ denotes the $i$th honest trainer's total contribution and $L'$ out of the total $L$ trainers are honest.

In addition, since the aggregator has performed per round validation accuracy computations throughout training (for hyperparameter tuning), it sends an additional proof for these computations to the smart contract. This proof is also based on the ZKP for accuracy evaluation described in Section~\ref{subsection:ZKP_ML}.

\noindent{\textbf{Verification and reward spreading.}} If the contribution assessment and validation accuracy computations are verified, the rewards are distributed from the project owner to the participating honest trainers based on the contribution of each trainer. In addition, the project owner compensates the trainers based on their participation. That is, even if an honest trainer does not improve the model, that trainer is still rewarded for participation. The reward of honest trainer $i$ is then given by 
\begin{align}
R_i = R_p + \alpha \frac{\bar{C}_i}{\sum_{j \in [L']} \bar{C}_j},
\end{align}
where $R_p$ is the fixed reward for participation, $\alpha$ is a parameter that depends on the total reward to be given (based on the accuracy improvement of the model), and $\bar{C} \triangleq \max(0,C)$. Here, we use $\bar{C}_i$ instead of $C_i$ since trainers with negative final contribution would not get any reward other than fixed $R_p$. Malicious trainers are not rewarded. As specified in the smart contract, based on the final accuracy of the trained model on test data, the project owner may get back some of the tokens he/she deposited in the beginning in case the model is not sufficiently improved by the trainers.

In addition to the participating honest trainers, the aggregator also receives monetary compensation for its operations during the execution of the marketplace. The aggregator is operated by an independent MLOps platform and paid separately by the project owner, according to an agreement between the project owner and the aggregator. 

\begin{Remark}
    In our design, validation and test datasets are provided by the project owner. These two datasets determine the final accuracy of the global model as well as the individual trainer contributions. A cheating project owner may purposefully send a biased or imbalanced test dataset (that is not similar to the validation dataset) to pay less to the trainers.\footnote{The project owner is not motivated to send an inaccurate validation dataset as this would hurt the training and yield a worse final model.} To solve this issue, we have the aggregator randomly partition the project owner's data. For this, in the beginning of the data market session, the project owner puts its data to IPFS and the corresponding CIDs are stored on the smart contract. Instead of letting the project owner specify the validation and test datasets, this time, we use a random seed generated by the trainers (through joint random number sharing algorithm \cite{bar1989non} or random number generation algorithm on blockchain \cite{du2019blockchain, simicc2020review}) in partitioning project owner's data into test and validation datasets so that the validation and test datasets are of similar (if not the same) distributions. 
\end{Remark}

\begin{Remark}
    In our design, the smart contract used to execute the entire data market operation is deployed by the project owner in the beginning. In order to avoid a project owner deploying a malicious smart contract, we utilize a factory contract that project owners can use to instantiate contracts for their projects. As long as the factory contract functions as intended, smart contracts it initiates should behave as expected. 
\end{Remark}

\begin{figure}[t]
	\centering
\includegraphics[width=0.49\textwidth]{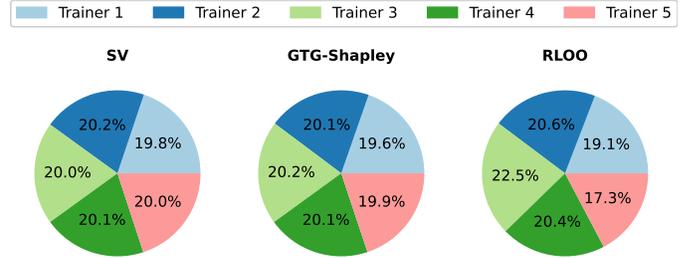}
	\caption{Trainer rewards under the i.i.d. scenario.}
	\label{fig:poc-iid}
    \vspace{-3mm}
\end{figure}

\begin{figure}[t]
	\centering
\includegraphics[width=0.49\textwidth]{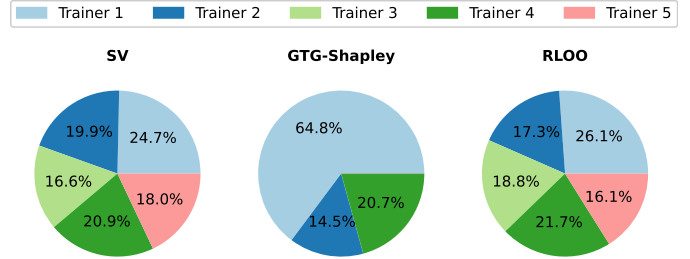}
	\caption{Trainer rewards under the non-i.i.d. scenario.}
	\label{fig:poc-noniid}
    \vspace{-5mm}
\end{figure}

\section{Experiments}
We provide experiment results to demonstrate the effectiveness/feasibility of the building blocks of the proposed design. In the outlier detection and contribution assessment experiments, we simulate the trainers and the aggregator on our in-house 64-bit Ubuntu 20.04.2 LTS machine equipped with AMD EPYC 7502 CPU. 
\subsection{Contribution Assessment}
\begin{figure}[t]
	\centering
\includegraphics[width=0.49\textwidth]{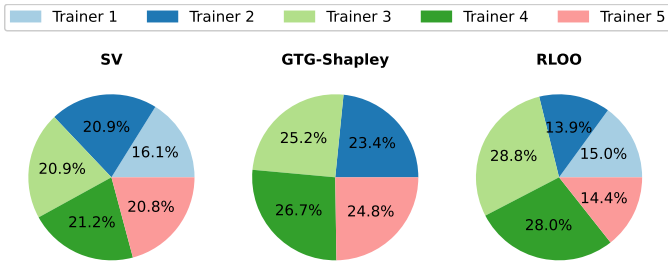}
	\caption{Trainer rewards under a rare data scenario with only Trainer 1 providing labels 0 and 1.}
	\label{fig:poc-noniid-rare}
    \vspace{-5mm}
\end{figure}

To demonstrate the performance of the proposed RLOO metric, we use the same setup as in GTG-Shapley work \cite{Liu2022gtg} and train a multi-layer perceptron (MLP) model on MNIST dataset, that is comprised of handwritten digits. We consider 1) i.i.d. distribution where each trainer has the same distribution and the same number samples; 2) a highly non-i.i.d. scenario where each trainer still has the same number samples but the digits (labels) in each trainer are mutually exclusive, i.e., each label is in only one data owner; and 3) one of the trainers (Trainer 1 in our case) has rate data (labels 0 and 1) while all the other trainers have i.i.d. distributions from the remaining labels (labels 2-9). For all these cases, we plot $\bar{C}_i /(\sum_{j \in [L']} \bar{C}_j)$ with $L'=L=5$ in Figs.~\ref{fig:poc-iid}-\ref{fig:poc-noniid-rare}. Here, we want to highlight that unlike the GTG-Shapley approach \cite{Liu2022gtg}, the proposed RLOO method is not meant to approximate the SV. We propose RLOO as an alternative more efficient metric for ZKP generation purposes. In the i.i.d. case in Fig.~\ref{fig:poc-iid} we see that the GTG-Shapley method closely approximates the SV and the proposed RLOO yields consistent results where each trainer has similar contribution values as expected. On the other hand, in the non-i.i.d. case Fig.~\ref{fig:poc-noniid}, the GTG-Shapley can no longer approximate the SV as closely and yields negative contribution values for trainers 3 and 5 (i.e., these trainers receive zero reward under GTG-Shapley) while the results produced by RLOO are more consistent. In Fig~\ref{fig:poc-noniid-rare}, we provide the results for a non-i.i.d. distribution where only Trainer 1 has labels 0 and 1. Unlike the GTG-Shapley scheme, RLOO and SV assign positive contributions (albeit less from the rest) to Trainer 1 which has rare data. GTG-Shapley scheme fails to acknowledge the contribution of Trainer 1. The key takeaway from these three experiments is that there is no universal way to accurately and effectively evaluate the trainer contributions in collaborative ML yet but the proposed data market in this work is compatible with various existing approaches to trainer valuation in collaborative ML.

\subsection{Outlier Detection}
We set the cosine similarity threshold $\gamma$ to 0.3 in the cross-round check. That is, if the cosine similarity score in (\ref{cosinesim}) is lower than 0.3, the aggregator deduces that there is an attack in the system.  
We consider two attacks: 1) Byzantine attack that produces random models for malicious trainers and 2) backdoor attack that scales up the weights of a malicious model
to dominate the global model~\cite{bagdasaryan2020backdoor}. In outlier detection experiments, we train a logistic regression (LR) model on the MNIST dataset and a more complex ResNet-56 model on the CIFAR10 dataset for both attack types using $8$ trainers for a total of 10 rounds. We set the number of malicious trainers to $3$ in the Byzantine attack scenario and $1$ in the backdoor attack case. We present the accuracy of our approach to successfully detect attacks (cross-round accuracy), as well as the accuracy of correctly removing the poisoned models (cross-trainer accuracy) in Table~\ref{tab:outlier_detection_exp}. Here, we would like to highlight that the accuracy values listed in Table~\ref{tab:outlier_detection_exp} are not test accuracies of the model but the accuracies of detecting and removing outliers, respectively. A 100\% accuracy means that attacks are detected and all outliers are removed. The experiment results provided in Table~\ref{tab:outlier_detection_exp} demonstrate the effectiveness and potential of the proposed outlier detection approach in collaborative ML.

\begin{table}[h]
\caption{Experimental results for the Outlier Detection}
\label{tab:outlier_detection_exp}
\centering
\begin{tabular}{ |c|c|c|c| } 
\hline
Attack & Model & Cross-round accuracy & Cross-trainer accuracy\\\hline
Byzantine & LR &	100\% &	100\%\\\hline
Byzantine & Resnet56 &	100\% &	100\%\\\hline
Backdoor & LR &	100\% &	90\%\\\hline
Backdoor & Resnet56 &	100\% &	100\% \\\hline
\end{tabular}
\vspace{-5mm}
\end{table}

\subsection{ZKP Performance}
We implement the ZKP system in Rust using the Arkworks framework \cite{arkworks}. The system consists of a machine learning module where various layers such as linear, embedding, and ReLU are implemented, and also an integration module that contains the outlier detection and the contribution assessment blocks. We test our system on Amazon AWS, using an m5a.4xlarge instance with 16 CPU cores and 32 GB memory.

We first test the inference of different models. In our experiments, we include LR and MLP models along with the Deep Learning Recommendation Model (DLRM). We sample a batch of 100 data points from the MNIST dataset and the Criteo Terabyte Dataset as our test set. We report the performance of a single model inference in Table~\ref{tab: zkp_model_inference}. We note that the proving time ranges from 4 seconds to 400 seconds depending on the complexity of the model, but the verification time remains the same thanks to the Groth16 \cite{groth2016size} proving system, which helps the smart contract deployment.

\begin{table}[h]
\caption{Model Inference Cost In ZKP}
\label{tab: zkp_model_inference}
\centering
\begin{tabular}{ |c|c|c|c| } 
\hline
Model & Circuit Size & Proving (s) & Verification (ms)\\\hline
LR & 230,440 & 4.21 & 3 \\\hline
MLP & 1,629,932 & 22.375 & 3 \\\hline
DLRM & 31,860,124 & 401 & 3 \\\hline
\end{tabular}
\end{table}

\begin{table}[h]
\caption{Cost of the Entire ZKP System. $^*$ denotes estimation.}
\label{tab: zkp_whole_system}
\centering
\begin{tabular}{ |c|c|c|c| } 
\hline
System & Circuit Size & Proving (s) & Verification (ms)\\\hline
LR & 5,605,226 & 75.364  & 3 \\\hline
MLP & 43,974,762 & 591 & 3 \\\hline
DLRM & 661,118,783 & 2.5h$^*$ & 3 \\\hline
\end{tabular}
\end{table}

\begin{figure}[t]
	\centering
        \includegraphics[width=0.49\textwidth]{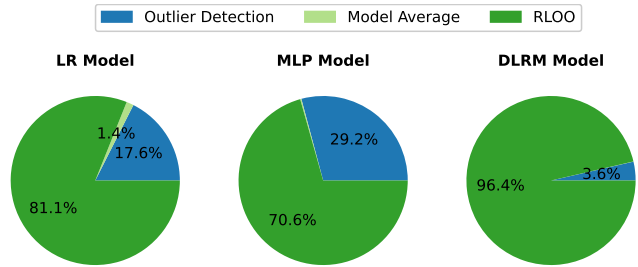}
	\caption{Percentage of Different Components in ZKP.}
	\label{fig:zkp_components}
    \vspace{-3mm}
\end{figure}

After that, we integrate the model into the entire system with outlier detection, model aggregation, and contribution assessment using RLOO. We report the performance in Table~\ref{tab: zkp_whole_system}. For the collaborative training of an LR model, we manage to prove its correctness in fewer than 1.5 minutes, while for the most complex model, i.e., DLRM, proving will take roughly 2.5 hours. Verification time still remains constant.

While the cost of each part differs depending on the data and model, we note that in general, the bottleneck of the system comes from the contribution assessment part, i.e., RLOO. It requires multiple times of model aggregation and model inference, which are extremely costly in practice. We show the percentage of circuit size in Fig.~\ref{fig:zkp_components}. Optimizing the circuit size for ML models is an open problem in the ZKP research community, and we expect further improvement with the progress of the research.

\section{Conclusion}
We proposed a trustworthy proof-of-contribution-based data marketplace design for collaborative/federated learning on blockchain. Key features of our design include contribution-based fair reward allocation, data privacy, robustness to malicious activities, verifiability through ZKPs, and universal applicability. We proposed novel outlier detection and proof-of-contribution techniques as well as a proof-of-concept ZKP implementation as building blocks. Our data market design is modular in that we can switch to other existing schemes in these building blocks and implement the data market with different trade-offs. Current focus is on the end-to-end deployment of the proposed design.

\bibliographystyle{unsrt}
\bibliography{IEEEabrv, References}
%



\end{document}